\documentclass[a4paper]{article}

\usepackage{icrc2013}
\usepackage[english]{babel} 

\title{The ANDES Deep Underground Laboratory}

\shorttitle{The ANDES Deep Underground Laboratory}

\authors{
X. Bertou$^{1}$
}

\afiliations{
$^1$ Centro At\'omico Bariloche, CNEA/CONICET
}

\email{bertou@cab.cnea.gov.ar}

\abstract{
ANDES (Agua Negra Deep Experiment Site) is a unique opportunity to build
a deep underground laboratory in the southern hemisphere. It will be
built in the Agua Negra tunnel planned between Argentina and Chile, and
operated by the CLES, a Latin American consortium. With 1750m of rock
overburden, and no close-by nuclear power plant, it will provide an
extremely radiation quiet environment for neutrino and dark matter
experiments. In particular, its location in the southern hemisphere
should play a major role in understanding dark matter modulation signals.
}

\keywords{underground laboratory, dark matter search, neutrino physics}

\begin{document}
\maketitle


\section{Introduction}
\label{intro}

Deep underground laboratories are one of the main actors of
astroparticles physics nowadays. In the past half-century they provided
unique ways of studying weak interactive particles. Super-Kamiokande can
be taken as a perfect example of the importance of deep underground
experiments and their impact both in astrophysics and particle physics,
hence of their role in astroparticle physics. Today, a dozen deep
underground laboratories are running or under construction, offering
experimental area at up to 2300\,m below ground to run neutrino physics
studies, dark matter search, or low radioactivity background
measurements. All of them are located in the northern hemisphere, with
the southernmost one being the planned INO, close to 10$^\circ$N.

In the past, some temporary laboratories have been set up in the southern
hemisphere, but none of them has been maintained. One of them, in a gold
mine in South Africa, contributed to the discovery of
atmospheric neutrinos in 1965\,\cite{southafrica} (together with
\cite{india}). Another experiment was looking for Dark Matter
oscillation signal in an iron mine in Argentina in
1995\,\cite{dmargentina}. Searches for a suitable mine in Brazil and
Chile have been performed in the past, but without success.

There is currently a growing demand for a southern hemisphere laboratory,
as more signals of Dark Matter appear. For years, DAMA/LIBRA has claimed
the observation of a yearly modulation of their signals and attributed it
to Dark Matter\,\cite{dama}. In the standard Cold Dark Matter models, the
ones most favoured by our current understanding of cosmological
measurements, our galaxy should have a halo of Weakly Interactive Massive
Particles (WIMP) in which we would be moving as the Sun moves through the
galaxy (at $\approx$232\,km/s). The movement of the Earth around the Sun
at 30\,km/s will make a modulation of the resulting WIMP wind. The
maximum signal is expected in June, when the Earth is moving towards
Cygnus, while the minimum is expected in December. DAMA/LIBRA observes
such a modulation at an 8.9\,$\sigma$ significance. However, it is
not clear if this effect is indeed a genuine Dark Matter effect or if it
could be some atmospheric or weather related effects. Different
interpretations are available in the literature\,\cite{muonsdama}.
Recently, more signals were reported by CoGeNT\,\cite{cogent} and
CDMS~II\,\cite{cdms}, and to get a clear confirmation of the modulation first
observed by DAMA, one would need an experiment in the southern hemisphere
observing the same modulation. The observation of an opposed modulation
would support the signal as coming from an atmospheric effect. As no
southern hemisphere laboratory is available, an experiment to the south
pole, DM-ICE\,\cite{dmice}, is being conducted, in a very hard to work
environment.

The news that a new road tunnel would be built in the Andes to link
Argentina and Chile was seen in 2010 as a unique opportunity to build a
deep underground laboratory in the southern hemisphere, in a similar way
the LNGS and LSM were built in Italy and France.

\section{The Agua Negra tunnel and the ANDES deep underground laboratory}

The Andes represent a natural barrier between Argentina and Chile in the
southernmost part of America. It has become of strategic importance for
Argentina and Brazil to be able to access the Asian market. The main
tunnel currently used to cross the Andes in Argentina is the Cristo
Redentor tunnel between Mendoza and Santiago de Chile. It can however
close during winter because of strong snows, and is not fully adapted to
the increasing commercial exchanges.

For years, various options have been proposed to complement the Cristo
Redentor tunnel. With the increasing political stability in South America
and the development of MERCOSUR and UNASUR, these projects were brought
back to improve the regional integration. The main project to improve the
connectivity between Argentina and Chile is the Agua Negra tunnel,
between San Juan and Coquimbo.

\begin{figure}[!ht]
\centering
\includegraphics[width=0.45\textwidth]{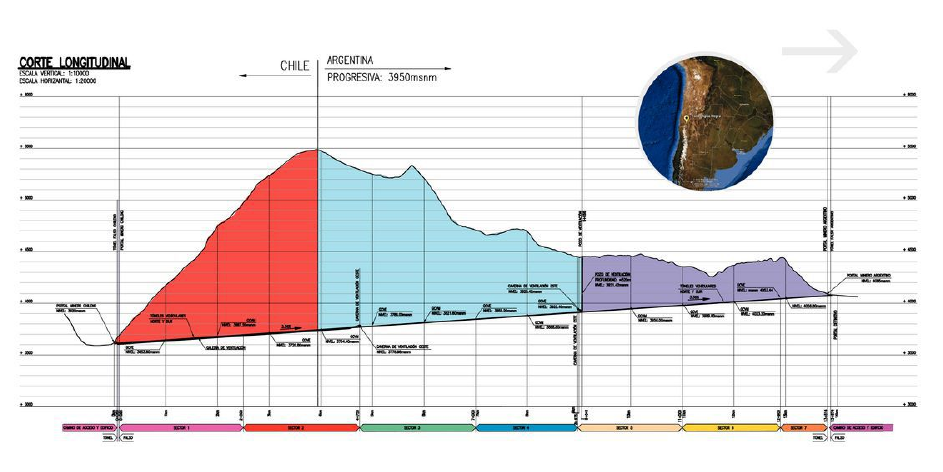}
\caption{Longitudinal cut of the Agua Negra tunnel and its location.}
\label{fig:tunelloc}
\end{figure}

A pre-feasibility study for the Agua Negra tunnel was finished in 2005,
and in 2008 a geological study campaign was started. In 2012, the final
project was proposed. In the meanwhile, the project was pushed forward
politically in numerous integration meetings in the region, such as the
treaty of Maipu (2009), the San Juan MERCOSUR meeting (2010), and in March
2012 the presidents of Argentina and Chile gave the green light for the
public tender. The public tender process started in January 2013, and the
international call for companies to compete in the tender was issued in
May 2013. The tender process is
expected to last for the whole 2013, given the difficulties of the
bi-national civil work. The construction of the tunnel should start in
2014, and last for 7 years.

The tunnel design consists in a double 14\,km long road tunnel, each 12\,m
of diameter, separated by 60\,m. The entry points are at high altitude,
4100\,m above sea level on the Argentine side, 3600\,m on the Chilean
side. Given the slope, most of the ventilation for the tunnel is natural,
with a forced ventilation system used in case of emergencies.
The deepest point of the tunnel is
located below the international limit between both countries, at about
4\,km from the Chilean entry. With 1750\,m of rock overburden, this spot
is ideal to host a deep underground laboratory, the ANDES deep
underground laboratory. ANDES can be read as an acronym for Agua Negra
Deep Experiment Site. It would be at a depth
equivalent to Modane, shallower only to
Jin-Ping and SNOLab.

Given the relatively high altitude of the laboratory and its
international location, two support laboratories are planned for ANDES.
There are no close-by towns to the tunnel. The Argentine laboratory is
expected to be in Rodeo, a small town at 60\,km from the tunnel entrance.
It will be the closest support laboratory and be mostly used for day to
day activities and the running of the experiments in ANDES. The Chilean
laboratory will be located in La Serena, at 180\,km of ANDES but in an
internationally connected city, with strong scientific presence (such as
the ESO for example). It will be mostly used for the preparation of the
installation of the experiments and their testing.

\begin{figure}[!ht]
\centering
\includegraphics[width=0.45\textwidth]{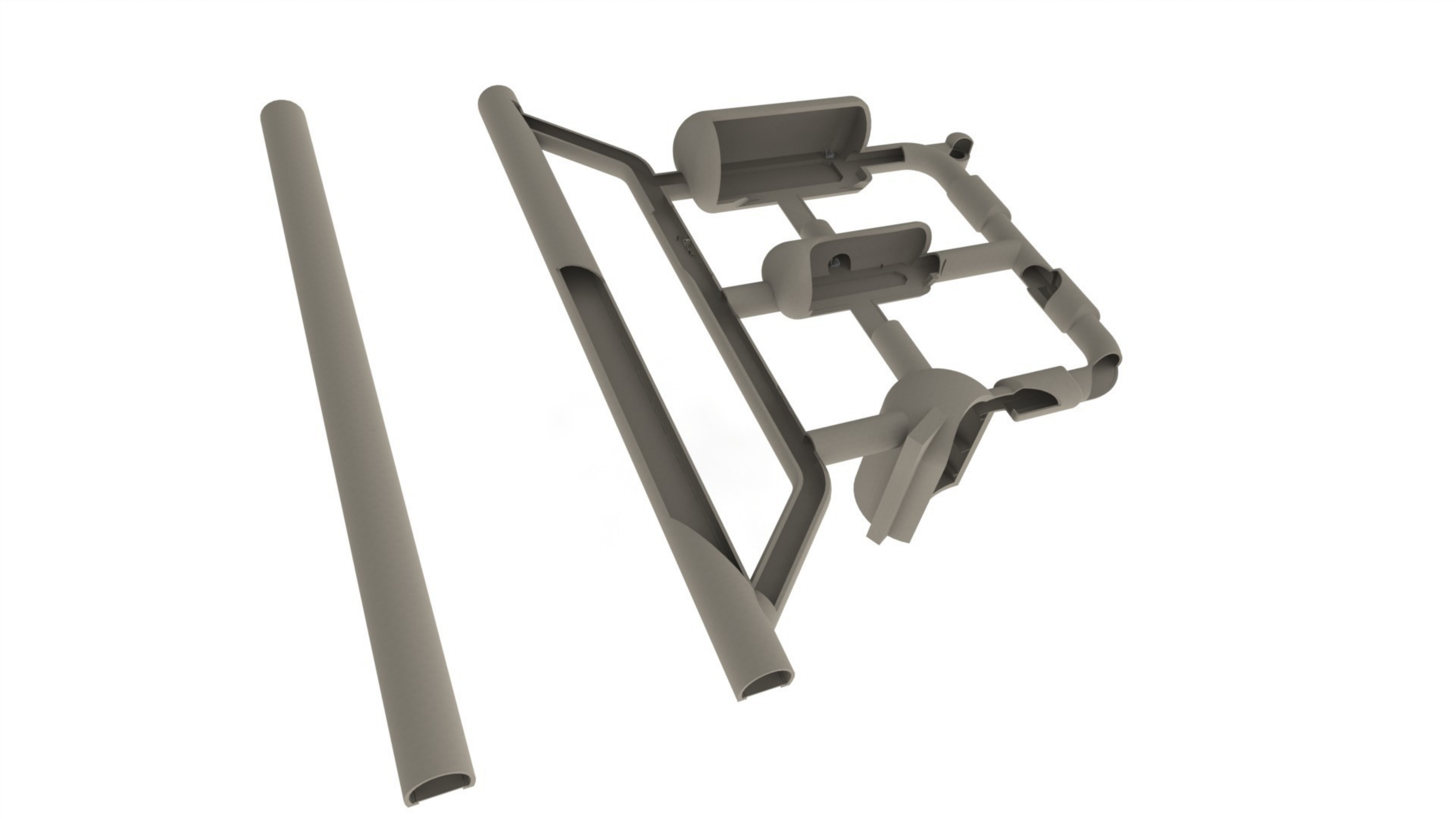}
\caption{Conceptual view of the ANDES deep underground laboratory,
located at km 4 of the Agua Negra tunnel.}
\label{fig:andes3d}
\end{figure}

The ANDES underground laboratory itself is foreseen to have a main hall of
21\,m width, 23\,m high and 50\,m long, to host large experiments
and a big pit of 30\,m of diameter and 30\,m of height for a single large
neutrino experiment. A
secondary cavern of 16\,m by 14\,m by 40\,m will host smaller experiments
and services, while three smaller caverns (9\,m by 6\,m by 15\,m) will
have dedicated experiments and a 9\,m diameter by 9\,m height pit will
focus on low radiation measurements. A conceptual layout of the
laboratory can be seen in Fig.\,\ref{fig:andes3d}.

\section{ANDES scientific programme}

The scientific programme of ANDES is similar to the one of any deep
underground laboratories, with some specificity due to its location.
The main topics in astroparticle physics are neutrino and dark matter.
There will be in addition a low radiation measurement laboratory, a
geophysics laboratory, space for biology experiments, and possibly a
particle accelerator to do nuclear astrophysics.

\begin{figure}[!ht]
\centering
\includegraphics[width=0.45\textwidth]{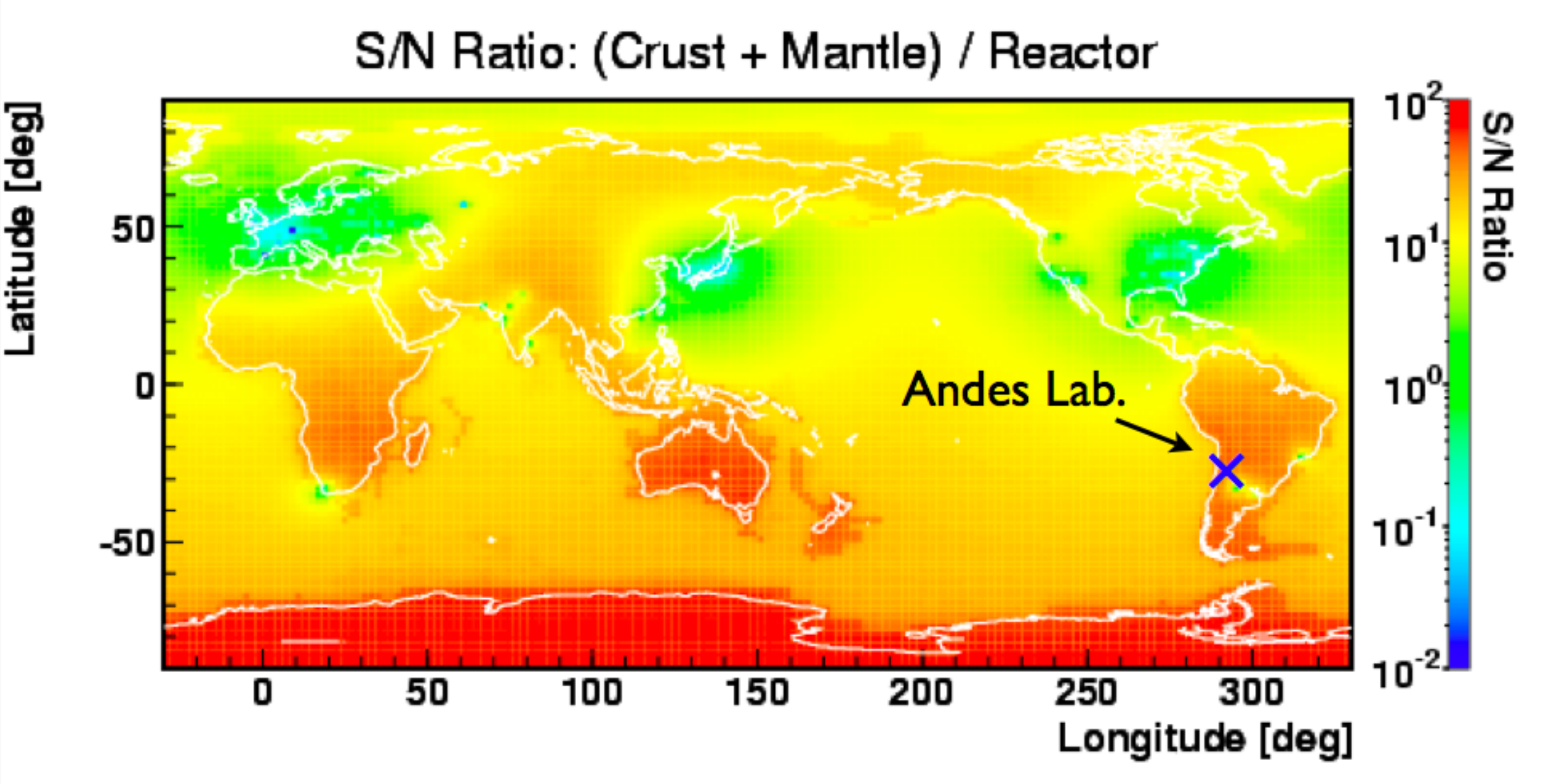}
\caption{Ratio of expected geo-neutrino signal to reactor background
over the Earth, taken from~\cite{andesgeo}. The nuclear power plants of
Argentina and Brazil are the only sources of background in South America,
and their influence does not reach the site of ANDES.}
\label{fig:andesgeo}
\end{figure}

In neutrino physics different experiments will be run. On one hand, ANDES
could host part of a large double beta decay experiment such as
SuperNEMO\,\cite{supernemo}. On the other hand, the flag experiment of
ANDES will be a large neutrino detector similar to
KamLAND\,\cite{kamland} and Borexino\,\cite{borexino}, but at a 3\,kton
scale\,\cite{renataneutrino}, focusing on low energy neutrinos. This
detector would allow complementary observation of neutrinos from a nearby
supernova, something essential to properly study the effect of matter on
neutrino oscillations. It will furthermore be an excellent geo-neutrino
observatory. Geo-neutrinos are produced in the Earth by radioactive decays
of Uranium, Thorium and Potassium. These decays are expected to be
responsible for a significant fraction of the Earth thermal balance, and
have been observed recently through their neutrino emission. When
detecting geo-neutrino, one has to face, in addition to the usual
background of any neutrino experiment, an extra background from
nuclear reactors. Current large neutrino detectors are in areas with
numerous nuclear reactors, and they have to deal with a large anti
electron neutrino background. ANDES is located in a nuclear quiet region,
far from the few reactors in Argentina and Brazil. The signal to noise
ratio is expected to be high, as can be seen on Fig.\,\ref{fig:andesgeo}.

As mentioned in the introduction, a critical reason to build ANDES is to
have southern hemisphere observation of dark matter modulation. While the
current signals reported by different experiments are controversial,
being incompatible with the results of Xenon 100\,\cite{xenon}, it is
likely that in the next 8 years while ANDES is built a genuine signal is
observed in a northern hemisphere deep underground laboratory. At that
stage, it will be logical to build a similar experiment in ANDES. In
addition, the important depth of ANDES will make it competitive for any
detection technique and it will likely host a third generation dark
matter search experiment.

An important part of the science programme will be dedicated to
geophysics. While it is not the main interest of the audience of this
conference, the detection of geo-neutrinos for example is a common
target for astroparticle physicists and geophysicists. It should be
mentioned that the Agua Negra tunnel is located in an area where there
is no volcanic activity, and earthquakes close to the surface, due to the
peculiar way the Nazca plate goes below the South American plate
(a Flat-Slab subduction\,\cite{gutscher}). This specific location makes
the laboratory very attractive to the geophysicist community.

\section{Radiation background expected at the ANDES deep underground
laboratory}

The radiation background was studied for the ANDES site, taken into
account the unusual high altitude and its possible impact on muon fluxes
and neutron activation.

\subsection{Muon flux}

The high energy muon flux (the one relevant for deep underground
laboratories, of $\approx$\,TeV energy scale) are not very dependant on the
altitude, given their high penetrating power. Low
energy muons fluxes and angular distribution start to change above 4\,km
of altitude, and some pions can be observed. All this has no relevance
after a few hundreds of metres of rock. The high geomagnetic cutoff of
the ANDES site does not either have any impact on the high energy muon
flux. These assumptions were verified with a set of complete cosmic ray
spectrum simulation based on CORSIKA\,\cite{corsika}.

\begin{figure}[!ht]
\centering
\includegraphics[width=0.45\textwidth,height=0.22\textwidth]{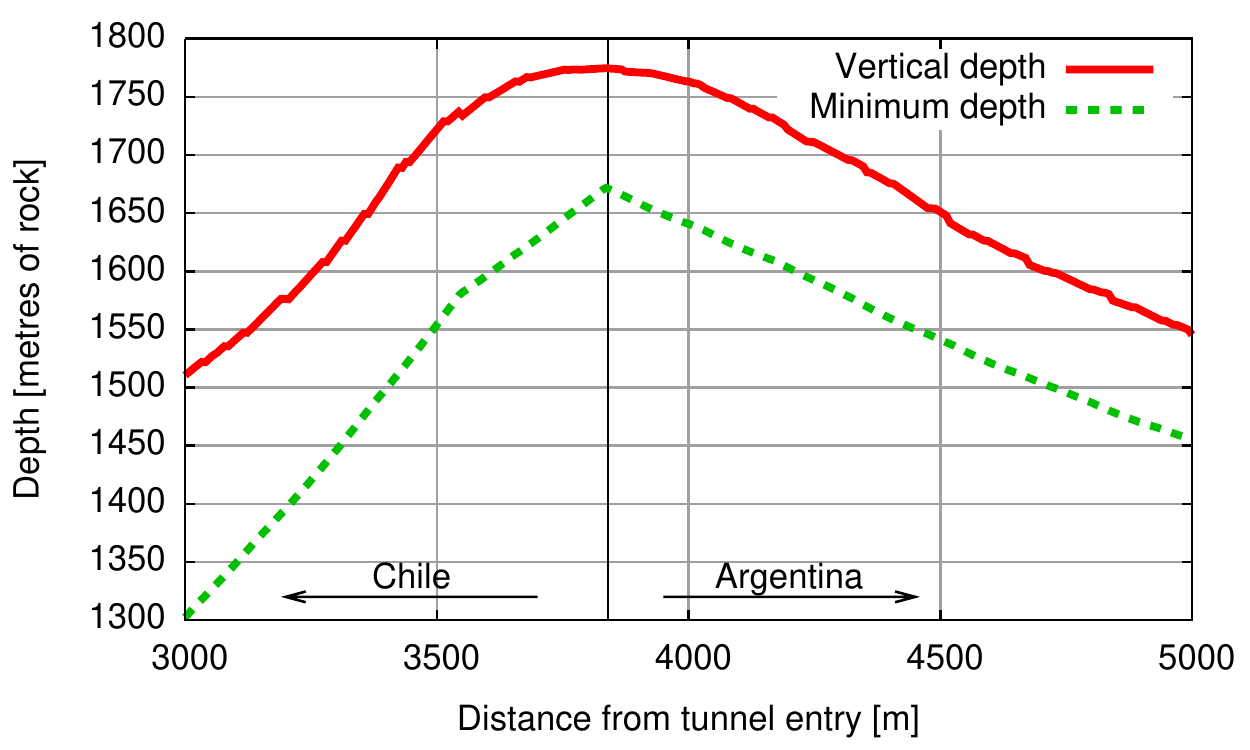}\\
\includegraphics[width=0.45\textwidth,height=0.22\textwidth]{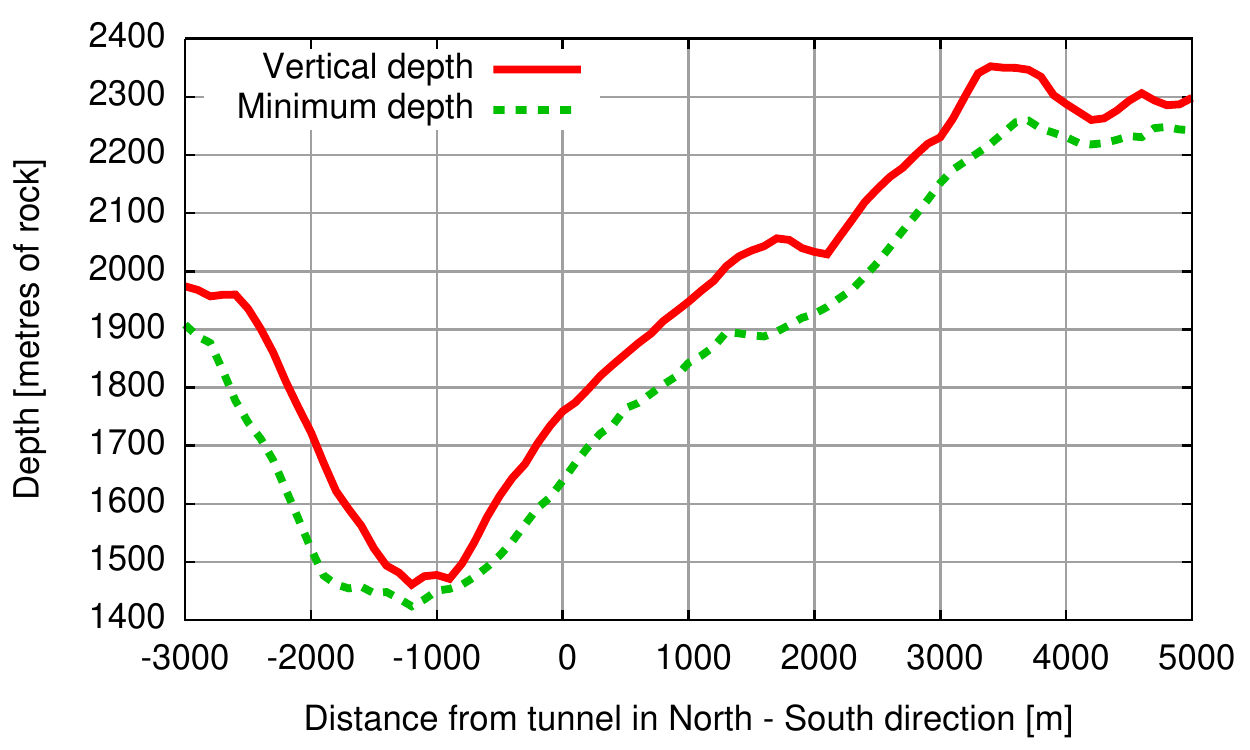}
\caption{Vertical and omnidirectional minimum depth obtained depending
on the location of the laboratory with respect of the tunnel. See text
for details.
}
\label{fig:tunelcut}
\end{figure}

To determine the rock overburden as a function of the precise location of
the laboratory along the Agua Negra tunnel, a preliminary layout of the
tunnel was used\,\cite{geoconsult} and the local geography was derived
from the Shuttle Radar Topography Mission\,\cite{nasaSRTM}. The vertical
and minimum omnidirectional depth were determined first considering the
laboratory to be 100\,m south of the tunnel axis, and then moving the
laboratory on the north-south axis around the deepest location.
The computed depths
can be seen in Fig.\,\ref{fig:tunelcut}, reaching a maximum of 1775\,m of
vertical depth and 1675\,m omnidirectional at 100\,m south of the tunnel
axis. More detailed geological studies are required to compute the water
equivalent depth of the laboratory, but given the expected average
density of 2.65\,g/cm$^2$, it is expected to be similar to
Modane.

\subsection{Rock radioactivity}

The precise geology at the deepest point of the tunnel is not currently
known and will not be determined with precision until the first
ventilation tunnel is dug. However, perforation of up to 650\,m deep were
performed in the exploratory phase and the deepest rock samples are
expected to be representative of what will be found 1000\,m deeper. The
main rocks present are Andesite, Basalt and Rhyolite (volcanic rocks).
Rhyolite is the most potential radioactive rock. Its content in potassium
in particular strongly depends on how it was formed.

4 samples were obtained from Geoconsult and analysed in the Neutron
Activation laboratory at the Bariloche Atomic Centre, measuring the
radiation lines from Uranium, Thorium,
and Potassium. All the samples were from deep extraction, between 450\,m and
650\,m deep. The results can be seen in table\,\ref{tab:rock}, compared
with measurements from Canfranc\,\cite{rockcanfranc}. Two samples of
rhyolite where analysed and both were found to have low content of
potassium. Given these results, the natural radioactivity expected for
the final location of the laboratory should be low.

\begin{center}
\begin{table*}[ht]
{\small
\hfill{}
\begin{tabular}{|l|l|c|c|c|c|c|c|c|}
\hline
& Basalt & Andesite & Rhyolite 1 & Rhyolite 2 & Canfranc\\
\hline
$^{238}$U & $2.6\pm0.5$ & $9.2\pm0.9$ & $14.7\pm2.0$ & $11.5\pm1.3$ & $4.5
- 30 $\\
$^{232}$Th & $0.94\pm0.09$ & $5.2\pm0.5$ & $4.5\pm0.4$ & $4.8\pm0.5$ &
$8.5 - 76 $\\
$^{40}$K & $50\pm3$ & $47\pm3$ & $57\pm3$ & $52\pm3$ & $37 - 880$\\
\hline
\end{tabular}}
\hfill{}
\caption{Radioactivity measurements of deep rock samples from the Agua Negra
tunnel. All values are in Bq/kg. Values for Canfranc are
from~\cite{rockcanfranc}.}
\label{tab:rock}
\end{table*}
\end{center}

\subsection{Neutron activation}

While it was shown that the high altitude of the laboratory was not an
issue for the muon flux, it is more relevant for the neutron atmospheric
flux responsible for neutron activation, lowering the purity of detector
material (not in the laboratory itself but in the support laboratories). In rare event experiments, radio-pure materials have to be
chosen, such as copper or archaeological lead. However, while copper has
no long lifetime isotope, it can be activated by neutrons into zinc, for
example
$^{65}$Zn, with an half life of 243.8 days. The activation rate is given
by the neutron flux, which is modulated by the altitude and the
geomagnetic latitude.

\begin{figure}[!ht]
\centering
\includegraphics[width=0.45\textwidth]{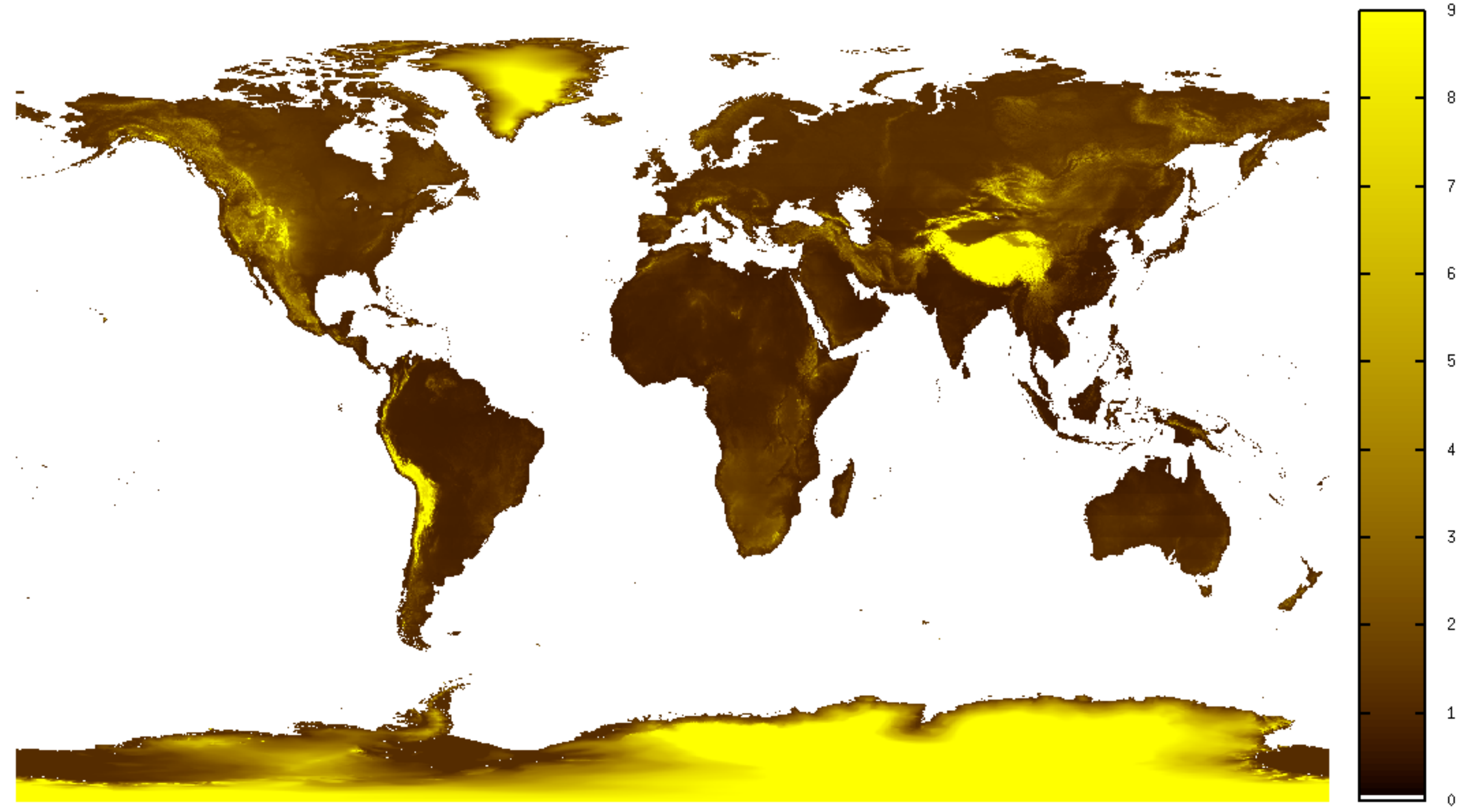}
\caption{Neutron flux at ground level relative to the average one at sea
level on the equator. The primary effect is due to the altitude, and a
secondary effect due to the geomagnetic latitude is visible.}
\label{fig:neutronmap}
\end{figure}

A simple model for the neutron flux can be obtained by multiplying two
factors, one given by the geomagnetic cutoff of the site, and one given
by the altitude\,\cite{neutronflux}. The geomagnetic cutoff average
values were computed on the globe using
Magnetocosmics\,\cite{magnetocos}, and the elevation map was taken from
ETOPO2\,\cite{etopo2}. The resulting neutron flux map can be shown on
Fig.\,\ref{fig:neutronmap}.

The neutron flux was determined at the two planned locations for the
support laboratories, Rodeo (1600\,m a.s.l.) and La Serena (sea level).
Given the proximity of the geomagnetic equator (which is displaced to the
south with respect of the geographic equator in the Americas), the final
neutron flux is lower than at other sites. The fluxes at Rodeo and La
Serena are found to be 2.2 and 0.9 (referred to the average at sea level
on the equator), while for example the fluxes at Modane and SNOLab are 2.3
and 1.4.

\section{The Latin American Consortium for Underground Experiments (CLES)}

ANDES was considered from the beginning as a unique opportunity not only
to build an underground laboratory for the international community but
to build directly an international laboratory. Given its location on the
borderline between two countries, and the current geopolitical unity
displayed by Latin American countries, ANDES was proposed to be run by a
consortium of Latin American countries, the CLES (initials of Latin American
Consortium for Underground Experiments in Spanish or Portuguese). The
CLES is currently formed by Argentina, Brazil, Chile and Mexico, and is
foreseen to open to more countries.

The CLES will be the organ in charge of the installation and operation of
the ANDES deep underground laboratory and its support laboratories. It
will also organise the academic integration of the scientific activities
in the laboratory with the regional systems. The CLES should be a pole
for underground science in the region.

\section{Conclusions and prospects}

The construction of the Agua Negra tunnel is a unique opportunity for the
construction of ANDES, the first deep underground laboratory in the southern
hemisphere. With the international tender process started in January
2013, the construction is expected to start in 2014, and the ANDES
laboratory could open in 2021.

ANDES is foreseen as an open international laboratory, coordinated by a
Latin American consortium (CLES). It will be hosting both
international experiments and regional ones. Its location will make it
an ideal laboratory to study dark matter modulation, supernovae neutrinos
and geo-neutrinos. Its large size and important depth will make it
competitive with all existing laboratories. In addition to a rich
astroparticle scientific programme, geophysics is expected to be a strong
part of the laboratory activity.

\section*{Acknowledgements}

This work has been possible thanks to
O.~Civitarese, C.~Dib, R.~Shellard and J.C.~d'Olivo, supporting the ANDES
initiative, the cosmic ray simulation and propagation data of H.~Asorey
and F.~S\'anchez, the radioactivity measurements of M.~Arribere and
M.~Gomez Berisso, the geomagnetic data of J.~Mas\'ias, the 3D layouts of
J.~Venturino, and Geoconsult Buenos Aires SA which provided the tunnel
layout and the rock samples.

This work has been supported by CNEA and CONICET, the
Argentine ministry for science and technology (MinCyT), and the San Juan province government.
%
%

\end{document}